\documentclass[manuscript,12pt]{article}
\usepackage{graphicx,color}
\usepackage{subfigure}
\usepackage{times}
\usepackage{mathptmx}
\usepackage{amsmath,amsthm,amssymb}
\makeatletter
\begin{document}
\title
{
\bf 
A possible mechanism of generation of extreme waves in shallow water
}

\author
{
Ken-ichi Maruno$^{1,2}$, Hidekazu Tsuji$^{3}$ and Masayuki Oikawa$^{3}$\\
\small
$^1$
Department of Mathematics, 
University of Texas-Pan American, Edinburg, TX 78541 USA\\
\small
$^2$
Faculty of Mathematics, 
Kyushu University, Hakozaki, Higashi-ku, Fukuoka, 812-8581, Japan\\
\small $^3$
Research Institute for Applied Mechanics,
Kyushu University, Kasuga, Fukuoka, 816-8580, Japan
}
\date{}

\maketitle
\abstract{
We consider a mechanism of generation of huge waves by 
multi-soliton resonant interactions. 
A non-stationary wave amplification phenomenon is found in 
some exact solutions of the Kadomtsev-Petviashvili (KP) equation. 
The mechanism proposed here explains the character
of extreme waves and of those in Tsunami.  
}
\section{Introduction}

The phenomenon of particularly 
high and steep waves on the sea surface is one of the most dangerous 
events. 
The large waves such as tsunami and freak (rogue) wave 
have a significant impact on the 
safety of people and infrastructure and are 
responsible for the erosion of coastlines and 
sea bottoms and the biological environment. 
For example, the Indian ocean tsunami of 26 December 2004 
caused an estimated 250,000 or more deaths and extensive 
damage due to run-up, landward inundation, and wave-structure 
interactions. 
The prediction of extreme waves is an important task for 
human being. 
Understanding the physics of large extreme waves may save 
lives. 

Freak waves occur much more frequently than it 
might be expected from surface wave statistics whereas they 
are particularly steep \cite{Kharif}. Freak waves may happen in both deep water
and shallow water. Although several physical mechanisms of generation of
freak waves in deep water have been proposed, the mechanism in shallow
water is few \cite{pelinovsky}. 

Recently, it has been established that a moving pressure 
disturbance can generate solitary waves also in open sea areas. 
This mechanism may occur in relatively shallow areas with 
heavy fast ferry traffic where soliton-like disturbances 
frequently occur. 
In a fatal accident which occured in Harwich, 
a port on England's east coast, in July 1999, 
one surviving victim reported that the 
wave looked like 'the white cliffs of Dover'. 
Research carried out in Europe shows that 
the soliton produced by a fast ferry was 
probably responsible for the disaster. 
This extreme wave is so-called 'solitary killer' wave 
\cite{Hamer,Li,Soomere-ferry}. 

Solitary waves in shallow water have been studied since the 19th century.
The theories of asymptotic expansion and integrable system 
have brought the remarkable progress to this research field.
With the assumption of weak nonlinearity the dynamics of solitary waves 
in shallow water is 
well described by the Korteweg-de Vries KdV) equation. 
Moreover, for weakly two-dimensional cases, that is, for the cases 
that the scale of variation in the direction normal to the propagation 
direction is much longer than that in the propagation direction, 
the fundamental equations may be reduced to the Kadomtsev-Petviashvili (KP) equnation. 
Both of the equations are integrable and it 
makes the precise analysis of the wave motion possible.

Peterson et al. studied the amplitude of two soliton solutions of 
the KP equation. 
It was pointed out that the interaction of two solitary waves may be one
of mechanism of generation of 
extreme waves \cite{Peterson,Soomere,Soomere2}.
In the case of resonant Y-shape
soliton which was found by Miles, the maximum amplitude of a
solitary wave can reach four times the amplitude of 
an incoming solitary wave \cite{Miles1, Miles2}. 
The weak point of this theory is that the interaction pattern of
2-soliton is stationary and does not describe non-stationary
phenomena. If a high wave is generated by soliton resonance once, the high
wave can not disappear within finite time.  
There is a gap to describe the generation of extreme wave using
2-soliton solutions of the KP equation（See \cite{Porubov}）.  

Recently, Biondini and Kodama studied a class of exact solutions of the
KP equation and found that the interaction patterns of solitons 
in the class have non-stationary web-like structures \cite{Gino}. 
Such solutions are resonant-type soliton solutions which are
the generalization of the Y-shape resonant soliton solution which was
found by Miles. 
These general resonant solutions are simply expressed by a Wronskian. 
Furthermore, Kodama gave a general formulation of the classification 
of interaction patterns of the $N$-soliton solutions of the KP equation 
\cite{Kodama} (See also \cite{Gino2}).

In this paper, we propose a mechanism of generation of huge waves by 
multi-soliton resonant interactions. The key of analysis is the
Wronskian solution of the KP equation.  

\section{Physical derivation of the Kadomtsev-Petviashvili equation}

Let us consider an object region of the ocean as an inviscid liquid layer of 
uniform undisturbed depth 
$h$ while the influence of the open air is assumed to be negligibly small. 
Assume the plane $z^*=0$ of the Cartesian coordinates 
$(x^*, y^*, z^*)$ coincides with 
the flat bottom of the region and 
$z^* = h + \zeta^*(x^*, y^*, t^*)$ denotes the free surface. 
Here, $t^*$ is time.
Let us denote the velocity components along the axes 
$x^*, y^*, z^*$ by 
$u^*(x^*, y^*, z^*,t^*)$, $v^*(x^*, y^*, z^*,t^*)$ and 
$w^*(x^*, y^*, z^*, t^*)$, respectively. 
Assume the fluid motion is irrotational, then the fluid velocity components are 
written as $u^* = \Phi^*_{x^*}, v^* = \Phi^*_{y^*}$, and $w^* = \Phi^*_{z^*}$ 
in terms of the velocity potential $\Phi^*(x^*, y^*, z^*, t^*)$. 
Here we consider the long waves of small but finite amplitude. 
Hence, we make the nondimensinalization as follows: 
\begin{equation}
(x^*, y^*) = \ell (x, y), \ z^* = hz, \ t^* = \frac{\ell}{c_0}t, \ \zeta^* = a\zeta, 
\ \Phi^* = \frac{a\ell c_0}{h}\Phi,
\label{nondimensionalization}
\end{equation}
where $\ell$ is the characteristic horizontal length, $a$ the 
characteritic amplitude, $g$ the gravitational accelaration, and 
$c_0 \equiv \sqrt{gh}$ the linear long-wave phase speed. 
Then the governing equations and the boundary conditions are: 
\begin{eqnarray}
&&\beta \Delta \Phi+\Phi_{zz}=0\,,\quad 
\quad \quad \quad \quad \quad \quad \quad  (0<z<1+\alpha \zeta)\,,\label{Laplace}\\ 
&&\Phi_z=0\,,\quad \quad \quad \quad 
\quad \quad \quad \quad \quad \quad \quad \quad (z=0)\,,\label{bottomBC}\\
&& \Phi_t+\frac{\alpha}{2}(\nabla \Phi)^2+
\frac{\alpha}{2\beta}\Phi_z^2+\zeta=0\,,
\quad \quad (z=1+\alpha \zeta)\,,\label{bound1}\\
&&\zeta_t+\alpha \nabla \Phi \cdot \nabla \zeta-\frac{1}{\beta}\Phi_z=0\,,
\quad \quad \quad \quad (z=1+\alpha \zeta)\,,\label{bound2}
\end{eqnarray}
where the subscript variables denote the partial differentiations 
with respect to the variables and 
$\nabla\equiv (\partial/\partial x,\partial/\partial y)$, 
$\Delta \equiv \partial^2/\partial x^2 + \partial^2/\partial y^2$, 
$\alpha=a/h$, and $\beta=(h/\ell)^2$. 
Here we consider weakly nonlinear long waves where weak nonlinearity and
weak dispersion balance out, thus we assume $\alpha \ll 1, 
\beta \ll 1$ and $\beta = O(\alpha)$ 
(hereafter we assume $\epsilon=\alpha=\beta$, that is, $\epsilon = a/h$ and 
$\ell = h/\sqrt{\epsilon}$). 
Then, from (\ref{Laplace}) and (\ref{bottomBC}), 
the velocity potential $\Phi(x,y,z,t)$ is represented as follows:  
\begin{equation}
\Phi(x,y,z,t)=\sum_{m=0}^{\infty}(-\epsilon \Delta)^m \phi(x,y,t)
\frac{z^{2m}}{(2m)!} \,,\label{expansion}
\end{equation}
where $\phi(x,y,t)$ is the value of the velocity potential at the bottom. 
Substituting (\ref{expansion}) into the boundary conditions
(\ref{bound1}) and (\ref{bound2}), we obtain 
the coupled equations
for the functions $\phi$ and $\zeta$
\begin{eqnarray}
&&\phi_t+\zeta+\frac{1}{2}\epsilon\nabla\phi\cdot\nabla\phi
-\frac{1}{2}\epsilon\Delta
 \phi_t=O(\epsilon^2)\,,\label{couple1}\\
&&\zeta_t+\Delta \phi+\epsilon\nabla\cdot(
\zeta\nabla\phi) 
-\frac{1}{6}\epsilon\Delta^2 \phi
=O(\epsilon^2)\,.\label{couple2}
\end{eqnarray}

For unidirectional propagation (in the positive $x$ direction) of waves, 
if we introduce the new variables 
$\xi = x - t, \ \sigma = \epsilon t$ and assume the expansions 
\begin{equation}
\zeta = \zeta_0(\xi, \sigma) + \epsilon\zeta_1(\xi, \sigma) + \cdots, \ \ 
\phi = \phi_0(\xi, \sigma) + \epsilon\phi_1(\xi, \sigma) + \cdots,
\end{equation}
we obtain the KdV equation for $\zeta_0$ :
\begin{equation}
2\frac{\partial \zeta_0}{\partial \sigma} + 3\zeta_0\frac{\partial \zeta_0}{\partial \xi} 
+ \frac{1}{3}\frac{\partial^3\zeta_0}{\partial \xi^3} = 0, \ \ 
\frac{\partial \phi_0}{\partial \xi} = \zeta_0.
\end{equation}
As is well known, this equation describes the dynamics of one dimensional 
soliton (line soliton). Miles formulated the interaction of two solitons 
propagaing in different directions ${\bf n}_1$ and ${\bf n}_2$ by means of 
a perturbation method \cite{Miles1}. Miles showed the followings: If the parameter 
$\kappa = \frac{1}{2}(1-{\bf n}_1\cdot{\bf n}_2) = \sin^2 \psi$ ($2\psi$ is 
the angle between ${\bf n}_1$ and ${\bf n}_2$) is not small, in the lowest 
order of approximation $\zeta$ can be represented by the superposition 
of two solitons propagating in the directions ${\bf n}_1$ and ${\bf n}_2$ 
which are governed by the respective KdV equations. But, if $\kappa = O(\epsilon)$, 
the perturbation expansion breaks down. Miles called the former the weak interaction 
and the latter the strong interaction.


In the case of $\kappa=O(\epsilon)$, the directions of propagation of two
solitons are almost the same. If we choose the $x$ axis 
as the main propagation direction, the scale of variation of the wave field in the 
$y$ direction would be $O(\epsilon^{-1/2})$. Thus, introducing the variables
\begin{equation}
\xi = x - t, \ \ \eta = \epsilon^{1/2}y, \ \ \sigma = \epsilon t, 
\label{scaled variavles}
\end{equation}
and expanding $\zeta$ and $\phi$ as
\begin{equation}
\zeta = \zeta_0(\xi, \eta, \sigma) + \epsilon \zeta_1(\xi, \eta, \sigma) + \cdots, \ \ 
\phi = \phi_0(\xi, \eta, \sigma) + \epsilon \phi_1(\xi, \eta, \sigma) + \cdots,
\label{expansion 2}
\end{equation}
we obtain the KP equation \cite{kp,Freeman}
\begin{equation}
2\frac{\partial \zeta_0}{\partial \sigma} + 3\zeta_0\frac{\partial \zeta_0}{\partial \xi} 
+ \frac{1}{3}\frac{\partial^3 \zeta_0}{\partial \xi^3} + 
\frac{\partial^2 \phi_0}{\partial \eta^2} = 0, \ \ \zeta_0 = 
\frac{\partial \phi_0}{\partial \xi}
\label{KP integral form}
\end{equation}
or 
\begin{equation}
\frac{\partial}{\partial \xi}\left(2\frac{\partial \zeta_0}{\partial \sigma} + 
3\zeta_0\frac{\partial \zeta_0}{\partial \xi} 
+ \frac{1}{3}\frac{\partial^3 \zeta_0}{\partial \xi^3}\right) + 
\frac{\partial^2 \zeta_0}{\partial \eta^2} = 0. 
\label{KP differential form}
\end{equation}
In terms of the dimensional variables, this is written as 
\begin{equation}
\frac{\partial}{\partial x^{*}}\left (\frac{\partial \zeta_0^{*}}{\partial t^{*}}
+ c_0\frac{\partial \zeta_0^{*}}{\partial x^{*}}
+ \frac{3}{2}\frac{c_0}{h}\zeta_0^{*} \frac{\partial \zeta_0^{*}}{\partial x^{*}}
+ \frac{1}{6}c_0h^2\frac{\partial^3 \zeta_0^{*}}{\partial x^{*3}}\right )
 + \frac{c_0}{2}\frac{\partial^2 \zeta_0^{*}}{\partial y^{*^2}}=0,
\label{KP dimensional form} 
\end{equation} 
where $\zeta_0^{*} = a\zeta_0$. 

Here, if we introduce the following dimensionless variables: 
\begin{equation}
u = \frac{3\zeta_0^{*}}{2h}, \ \ X = \frac{1}{h}(x^{*} - c_0 t^{*}), \ \ 
Y = \frac{y^{*}}{h}, \ \ T = -\frac{2c_0}{3h}t^{*}, 
\end{equation} 
then (\ref{KP dimensional form}) becomes 
\begin{equation}
(-4u_T + u_{XXX} + 6uu_X)_X + 3u_{YY} = 0.
\label{KP-5}
\end{equation}
Further, introducing $\tau$-function through
\begin{equation}
u = \frac{3\zeta_0^{*}}{2h} = 2(\log \tau)_{XX}, 
\label{tau-function}
\end{equation}
we obtain the following bilinear form from (\ref{KP-5}):
\begin{equation}
(D_X(-4D_T + D_X^3) + 3D_Y^2)\tau\cdot\tau = 0, 
\label{KPbilinear}
\end{equation}
where $D_X$, $D_Y$ and $D_T$ are the Hirota bilinear operators defined, for example, by 
\[
 D_x^nD_y^mf\cdot g=\left(\partial_x-\partial_{x'}\right)^n
\left(\partial_y-\partial_{y'}\right)^mf(x,x')g(y,y')|_{x'=x,y'=y}\,. 
\]
Using the Hirota bilinear method \cite{Hirota}, 
the $\tau$-function of 1-soliton solution is given by 
\begin{eqnarray}
&&\tau=1+\exp(2\varTheta),\quad \varTheta 
= {\bf K}\cdot {\bf X} + \varOmega T + \varTheta^{0}
= KX + LY + \varOmega T + \varTheta^{0}\,,\\
&&\varOmega = K\left(K^2 + \frac{3L^2}{4K^2}\right)\,, 
\end{eqnarray}
and then the dimensionless elevation $\tilde{\zeta} = \zeta_0^{*}/h$ is given by 
$\tilde{\zeta} = \frac{4}{3}(\log \tau)_{XX} = \frac{4K^2}{3}{\rm sech}^2 \varTheta$.
The $\tau$-function of 2-soliton solution is given by 
\begin{eqnarray}
&&\tau = 1 + \exp(2\varTheta_1) + \exp(2\varTheta_2) + A\exp(2\varTheta_1+2\varTheta_2)\,,\\
&&\varTheta_i = {\bf K_i}\cdot {\bf X} + \varOmega_i T + \varTheta_{i}^{0} 
= K_iX + L_iY + \varOmega_i T + \varTheta_{i}^{0}\,,\\
&&\varOmega_i = K_i\left(K_i^2 + \frac{3L_i^2}{4K_i^2}\right)\,,
\quad (i=1,2)\,,\\
&&A=\frac{4(K_1-K_2)^2-(\tan \psi_1-\tan \psi_2)^2}
{4(K_1+K_2)^2-(\tan \psi_1-\tan \psi_2)^2}\,, \\
&&\tan \psi_i=L_i/K_i,\quad (i=1,2)\,.
\end{eqnarray}
which is the most well-known 2-soliton solution of the KP equation 
\cite{satsuma}. 
In the next section, we study more general soliton solutions which are 
expressed by the Wronskian form. 

\section{Wronskian solutions, resonant-soliton and extreme elevations}

If the functions $f_i(X, Y, T)$ satisfy the relations
\begin{equation}
\frac{\partial f_i}{\partial Y}=
\frac{\partial^2 f_i}{\partial X^2},\,\quad 
\frac{\partial f_i}{\partial T}=
\frac{\partial^2 f_i}{\partial X^3}\,,\quad \quad (i=1,\cdots,N), 
\label{the relations} 
\end{equation}
then the Wronskians
\begin{equation}
\tau_N=
\left|\begin{array}{ccc}
f_1^{(0)} &\cdots &f_N^{(0)}\\
\vdots &\ddots &\vdots\\
f_{1}^{(N-1)} &\cdots &f_N^{(N-1)}
\end{array}\right|, \ \ \ 
\mbox{with} \ \ f_i^{(n)}=\frac{\partial^n f_i}{\partial X^n}\,,\quad (i=1,\cdots, N)
\label{kptau}
\end{equation}
%
%
are solutions to the bilinear equation (\ref{KPbilinear}) \cite{F-N}. 
For example, we can obtain the ordinary $N$-soliton solutions by setting
$f_i$ as
\begin{eqnarray}
&&f_i = e^{\theta_{2i-1}} + e^{\theta_{2i}}\,,\quad (i = 1,\cdots,N)\,,\label{f of soliton}\\
&&\theta_j = -k_jX + k_j^2Y - k_j^3T + \theta_j^0\,,\quad (j=1,\cdots,2N)\,,\label{phase of soliton}
\end{eqnarray}
where $k_j$ and $\theta_j^0$ are constants. 

Let us choose the following $f_i$ as $f_i$ satisfying (\ref{the relations})
\begin{eqnarray}
&&f_1=\sum_{i=1}^M e^{\theta_i}=:f\,,
\quad f_i=f^{(i-1)}\,,\quad 1<i\leq N\leq M, \,.\label{f-toda}
\end{eqnarray}
where $\theta_i, (i=1,2,\cdots,M)$ are given in the form (\ref{phase of soliton}) 
and the ordering $k_1 < k_2 < \cdots < k_M$ is assumed. 
Then $\tau$-function in (\ref{kptau}) takes the form 
\begin{equation}
\tau_N=
\left|\begin{array}{ccc}
f^{(0)} &\cdots &f^{(N-1)}\\
\vdots &\ddots &\vdots\\
f^{(N-1)} &\cdots &f^{(2N-2)}
\end{array}\right|.  \label{tau-toda}
\end{equation}
Using the Binet-Cauchy theorem, the $\tau$-function (\ref{tau-toda})
with (\ref{f-toda}) can be expanded as a sum of exponential functions, 
\begin{equation}
\displaystyle{
\tau_{M_+}=
  \kern-1em \sum_{1\le i_1<\cdots<i_{M_+}\le M}
    \Delta(i_1,\dots,i_{M_+})\,\,
\exp\bigg(\sum_{j=1}^{M_+}\theta_{i_j}\bigg)\,,}
\label{tau-resonant}
\end{equation}
where $M=M_++M_-$, $N=M_+, (1\leq M_+ \leq M-1)$. 
$\Delta(i_1,\dots,i_{M_+})$ is the square of 
Vandermonde's determinant, 
$\Delta(i_1,\dots,i_{M_+})=\prod_{1\le j<l\le M_+} 
(k_{i_j}-k_{i_l})^2$ 
\cite{Gino}. This is called the fully resonant $(M_-,M_+)$-soliton solution.
This solution has $M_-$ solitons in asymptotics for $y \to -\infty$,
$M_+$ solitons in asymptotics for $y \to \infty$, 
in the intermediate region these solitons interact resonantly. 
The $\tau$-function (\ref{tau-resonant}) 
is a general form of fully resonant-type soliton solutions. 

\begin{figure}[t!]
\begin{center}
\end{center}
\caption{Fully resonant soliton solutions}
\label{1-2-2-2-soliton}
\end{figure}

It is also possible to set $f_i$ as
\begin{equation}
\label{f-function}
f_i=\sum_{j=1}^M a_{ij}\,e^{\theta_j}\,,\quad {\rm for}\quad i=1,
\ldots,N,\quad
{\rm and}~~M>N\,,
\end{equation}
using the phases $\theta_j$ given in the form (\ref{phase of soliton}) and 
a $N\times M$-matrix $A_{(N,\,M)}:=(a_{ij})$. 
Then the $\tau$-function ($\ref{kptau}$) with (\ref{f-function}) 
is expanded as a sum of exponential functions,
\begin{equation}
\tau=\sum_{1\le i_1<\cdots<i_N\le M}\xi(i_1,\ldots,i_N)\prod_{1\le j<l\le N}
(k_{i_j}-k_{i_l})\exp\left(\sum_{j=1}^{N}\theta_{i_j}\right)\,,
\label{tau-general}
\end{equation}
where $\xi(i_1,\ldots,i_N)$ is the $N\times N$-minor whose $j$-th columns 
($j = 1, \cdots, N$) are 
given by the $i_j$-th columns in the matrix 
$A_{(N,\,M)}=(a_{ij})$, respectively, 
\[
\xi(i_1,\ldots,i_N):=
\left|
\begin{array}{ccc}
a_{1,\,i_1}&\cdots& a_{1,\,i_N}\\
\vdots &\ddots &\vdots \\
a_{N,\,i_1}&\cdots&a_{N,\,i_N}
\end{array}
\right|\,.
\]
This is a general form of solutions in which the line solitons interact
resonantly and nonresonantly \cite{Kodama}. The $\tau$-function 
(\ref{tau-resonant}) of the fully resonant line soliton solutions is a
special case of the general $\tau$-function (\ref{tau-general}). 
In this general form of line soliton solutions, 2-soliton solutions 
in the sense that the solutions have the same sets of two line solitons 
in both asymptoyics for $y \to \pm\infty$ 
are classified into three types, O-type, P-type and T-type. 
The $A$-matrices $A_{(2,4)}$ corresponding these three types have 
the following row reduced echelon forms
\[
 A_{(2,4)}^{(\rm O)}=\left(\begin{array}{cccc}
1 & 1 & 0 & 0 \cr
0 & 0 & 1 & 1 \cr
\end{array}
\right)
\,,\quad 
 A_{(2,4)}^{(\rm P)}=\left(\begin{array}{cccc}
1 & 0 & 0 & -1 \cr
0 & 1 & 1 & 0 \cr
\end{array}
\right)
\,,\quad 
 A_{(2,4)}^{(\rm T)}=\left(\begin{array}{cccc}
1 & 0 & - & - \cr
0 & 1 & + & + \cr
\end{array}
\right)\,,
\]
respectively, where `$+,-$' shows the signs of the entries (also
nonzero). T-type soliton is resonant two soliton solution, and O-type
soliton is corresponding to two soliton solution obtained by the Hirota
direct method \cite{Hirota}. 
Note that all $N$-soliton solutions 
in the sense that the solutions have the same sets of $N$ line solitons 
in both asymptoyics for $y \to \pm\infty$ are classified by using Young
diagrams, i.e. we can construct the $A_{(N,2N)}$-matrices from the Young
diagrams \cite{Kodama}. 
\begin{figure}[t!]
\begin{center}
\end{center}
\caption{The interaction pattern of the fully resonant $(2,3)$-soliton 
($k_1=-2,k_2=-1,k_3=-1/100,k_4=1/100,k_5=2$)}
\label{2-3-soliton}
\end{figure}

Biondini and Kodama \cite{Gino} and Kodama \cite{Kodama} considered 
the interaction patterns but did not the amplitudes of line solitons. 
Here, we consider the amplitudes generated by soliton interactions. 
First, we study fully resonant soliton solutions given by the
$\tau$-function (\ref{tau-resonant}). 
Figure \ref{1-2-2-2-soliton} shows three examples of resonant soliton solutions. 
Figure \ref{1-2-2-2-soliton}(a) shows the Y-shape resonant soliton solution 
(resonant $(2,1)$-soliton). 
The two incoming solitons interact resonantly and generate a high amplitude 
soliton. The amplitude of the high amplitude soliton generated by soliton resonance 
is four times as large as the amplitude of the incoming soliton. 
We see the amplitude of soliton is amplified by soliton resonance. 
Figures \ref{1-2-2-2-soliton}(b) and \ref{1-2-2-2-soliton}(c) show 
the resonant $(2,2)$-soliton (T-type 2 soliton). 
In general, T-type 2-soliton has a hole as seen in Fig.\ref{1-2-2-2-soliton}(b). 
However, in Fig.\ref{1-2-2-2-soliton}(c) we see no hole but H-shape 
soliton. 
This is a degenerate case of the T-type 2-soliton which occurs due to 
the closeness of two parameters $k_2$ and $k_3$. 
The amplitudes of the intermediate interaction region 
in Fig.\ref{1-2-2-2-soliton}(b) are not so high, i.e. smaller
than the double of the amplitude of the incoming soliton. 
On the other hand, in Fig.\ref{1-2-2-2-soliton}(c) 
we see the amplitude of the interaction region of the H-shape soliton 
is very near to four times the amplitude of the incoming soliton, 
i.e. higher than the sum of the amplitudes of two incoming solitons. 
In Fig.\ref{1-2-2-2-soliton}(b) the interaction region makes a hole,
and then the amplitude spreads out, then the amplitude becomes higher. 
In contrast, 
Fig.\ref{1-2-2-2-soliton}(c) has no hole and the amplitude
concenteates on small region. 
Physically, the existence of the H-shape soliton is predicted by Miles
by using symmetry argument of the Y-shape resonant soliton \cite{Miles2}. 
The H-shape soliton is not a degenerate solution of 2-soliton solution. 
The resonant 2-solitons having a hole have been found in numerical
simulations of the (2+1)-dimensional soliton equations \cite{yajima,tsuji}. 
It is natural that the H-shape solitons appear in physical situations
such as ocean surface waves. 
The Y-shape soliton is a limit of the H-shape soliton.

\begin{figure}[t!]
\begin{center}
\end{center}
\caption{The interaction pattern of the fully resonant $(3,3)$-soliton 
($k_1=-5/2,k_2=-5/4,k_3=-1/2,k_4=1/2,k_5=3/2,k_6=5/2$)}
\label{3-3-soliton}
\end{figure}

Figure \ref{2-3-soliton} shows the interaction pattern of fully resonant 
$(2,3)$-soliton. 
In contrast to ordinary nonresonant soliton solutions, 
the interaction pattern of fully resonant $(2,3)$-soliton 
shows complex nonstationary
behavior.  

Figure \ref{3-3-soliton} shows the interaction pattern of fully resonant 
$(3,3)$-soliton. 
We see there exist 4 holes. 
In general, resonant $(M_{-},M_{+})$-soliton solutions 
given by the $\tau$-function (\ref{tau-resonant}) generate web-like structure
having $(M_{-}-1)(M_{+}-1)$ holes \cite{Gino}. 

The existence of web structure 
makes us expect the existence of nonstationary solutions having 
maximum amplitude for finite time. Thus, we have a following question: 
are there
any exact solutions having characters of extreme waves? 
In Fig.\ref{2-3-soliton} and Fig.\ref{3-3-soliton}, 
we see that the maximum amplitude of fully 
resonant multi-soliton solutions decrease at certain time by contraries.   
Figure \ref{max} show the time evolution of the 
maximum amplitude of Fig.\ref{3-3-soliton}.
Fully resonant soliton solutions having web structure do not have 
characters of the extreme waves. 

\begin{figure}[t!]
\begin{center}
\end{center}
\caption{Time evolution of the maximum amplitude of
fully resonant $(3,3)$-soliton. Exact solution.} 
\label{max}
\end{figure}

\begin{figure}[t!]
\begin{center}
\end{center}
\caption{The time eveolution of the partial resonant 
soliton solution given by the 
$\tau$-function (\ref{tau-general}) with the A-matrix $A_{(2,5)}$
 (\ref{a25}), i.e., $2\times 2$-Wronskian, 
$f_1=\sum_{k=1}^{3}e^{\theta_k}$, $f_2=\sum_{k=4}^{5}e^{\theta_k}$　
($k_1=-2,k_2=0,k_3=2,k_4=2.01,k_5=4$)}
\label{3-2-soliton}
\end{figure}

\begin{figure}[t!]
\begin{center}
\end{center}
\caption{Plots of figure \ref{3-2-soliton} in the plane $y=$constant at 
$t=1$: (a) y=5.5, (b) y=4, (c) y=-2}
\label{amplitude}
\end{figure}

Next, we consider the $\tau$-function (\ref{tau-general}) which is the
general form of $N$-line soliton solutions of the KP equation.  
This solution includes rich variety of solutions and there are
nonstationary solutions which are similar to extreme wave. 
Figure \ref{3-2-soliton} shows an example.
The solution of Fig.\ref{3-2-soliton} was made by the 
$2\times2$-Wronskian solutions, the $A$-matrix is 
\begin{equation}
A_{(2,5)}=\left(\begin{array}{ccccc}
1 & 1 & 1 & 0 & 0 \cr
0 & 0 & 0 & 1 & 1 \cr
\end{array}
\right)\,. \label{a25}
\end{equation}
We see that the stem of Y-shape soliton 
and the other soliton interact near-resonantly. 
When 1-soliton cross with the stem of Y-shape soliton,  
the interaction region is amplified into high amplitude 
(See Fig.\ref{3-2-soliton}, Fig.\ref{amplitude} and Fig.\ref{bigwave}).  
Thus, if there are several resonant solitons and each
interaction parts of resonant solitons interact resonantly, 
huge waves are generated. 
Using this principle, we can easily understand that 
there is a mechanism of generation of the extreme waves. 

In Fig.\ref{inelastic-soliton}, we show the interaction of 
inelastic 2-soliton given by the $A_{(2,4)}$-matrix
\begin{equation}
A_{(2,4)}=\left(\begin{array}{cccc}
1 & 1 & 1 & 0 \cr
0 & 0 & 1 & 1 \cr
\end{array}
\right)\,. \label{a24}
\end{equation}
We see that the interaction pattern is a combination of two Y-shape
resonant solitons at $t<0$, i.e. there are two resonant stems.  
At $t=0$, two resonant stems have a head-on collision, 
then a big amplitude wave
is generated at $t>0$  (see Fig.\ref{amplitude2} and \ref{inelasticamp}). 

In Fig.\ref{cont}, we show the contour plots of solitons corresponding
to Fig.\ref{3-2-soliton} and Fig.\ref{inelastic-soliton}. 
Based on the asymptotic analysis which was discussed in \cite{Gino}, 
each solitons are labeled by the index $[i,j]$ which 
represents a soliton made by two dominant exponentials 
$e^{\theta_i}$ and $e^{\theta_j}$. 
Taking the limit $k_4\to k_3$, $k_5\to k_{4'}$
in the soliton solution of the case of $A_{(2,5)}$ 
in Fig.\ref{3-2-soliton}, we can recover the interaction pattern 
of the soliton solution of the case of $A_{(2,4)}$ 
in Fig.{\ref{inelastic-soliton}}. 
Because a new soliton resonance is made by the limit, the 
arrangement of solitons is changed.   
 
The mechanism of generation of extreme waves is as follows: 
(i) Several solitons are generated by some reasons, e.g. fast ferry
traffic, tsunami, current and topography. 
(ii) Several generated solitons interact resonantly. 
(iii) When the interaction region of a resonant soliton interacts with one of
the other resonant soliton, the amplitude will be 16 times of the
amplitude of a soliton. (iv) After the interaction of the
stem of resonant solitons, the amplitude decreases.  

This shows that the soliton resonance can be a mechanism of generation
of extreme waves. The resonant interaction of several solitary waves may 
cause extreme waves higher than 4 times of amplitude of a solitary
wave. 
This mechanism may happen in ocean; for instance, tsunami, freak wave
and solitary killer wave may be amplified by this mechanism. 

\section{Conclusion}

The KP equation has rich mathematical structure and many interesting
solutions. 
Finding physically interesting solutions of the KP equation is an
important problem. 
In this paper, we have searched solutions having characters of extreme
waves and found such solutions which are related resonant soliton
solutions expressed by the Wronskian form. 
We have found there exists the solutions similar to properties of 
the extreme waves on the sea surface. 
The mechanism of generation of extreme waves is, 
(i) several solitons are generated by some reasons, 
(ii) several generated solitons interact resonantly,  
(iii) when the interaction region of a resonant soliton interacts with one of
the other resonant soliton, the amplitude will be 16 times of the
amplitude of a soliton, (iv) after the interaction of the
stem of resonant solitons, the amplitude decreases.

We acknowledge helpful discussions with G. Biondini, S. Chakravarty and
Y. Kodama. 
KM also acknowledges support from the 21st Century 
COE program ``Development of Dynamic Mathematics with High Functionality'' 
at the Faculty of Mathematics, Kyushu University. 

\begin{figure}[t!]
\begin{center}
\end{center}
\caption{The change of maximum amplitude of figure \ref{3-2-soliton}.}
\label{bigwave}
\end{figure}

\begin{figure}[t!]
\begin{center}
\end{center}
\caption{The time eveolution of the inelastic 2-soliton solution given by the 
$\tau$-function (\ref{tau-general}) with the A-matrix $A_{(2,4)}$ (\ref{a24}) 
($k_1=-2,k_2=0,k_3=2,k_4=4$)}
\label{inelastic-soliton}
\end{figure}
\begin{figure}[t!]
\begin{center}
\end{center}
\caption{Plots of figure \ref{inelastic-soliton} in the plane $y=$constant at 
$t=1$: (a) y=5.5, (b) y=3.5, (c) y=-0.5}
\label{amplitude2}
\end{figure}
\begin{figure}[t!]
\begin{center}
\end{center}
\caption{The change of maximum amplitude of figure \ref{inelastic-soliton}.}
\label{inelasticamp}
\end{figure}
\begin{figure}[t!]
\caption{The contour lines of interaction patterns 
in Fig.\ref{3-2-soliton} and Fig.\ref{inelastic-soliton}. 
Each solitons are labeled by the index $[i,j]$ which 
represents a soliton made by two dominant exponentials 
$e^{\theta_i}$ and $e^{\theta_j}$. The upper graphs are corresponding to
 Fig.\ref{3-2-soliton}, the lower graphs are corresponding to
 Fig.\ref{inelastic-soliton}.}
\label{cont}
\end{figure}

\end{document}